\documentclass[a4paper,11pt]{article}
\usepackage{jcappub} % for details on the use of the package, please see the JINST-author-manual
\usepackage{soul}
\usepackage{lineno}
\usepackage{hyperref}
\usepackage{subcaption}

\newcommand{\cena}{Cen~A~}

\arxivnumber{}

\title{Centaurus~A: Opportunities to unveil particle acceleration with the CTAO}

\author{Cainã de Oliveira} %\note{Corresponding author.}
\author{and Edivaldo Moura Santos}
\affiliation{Instituto de Física, Universidade de São Paulo,\\
Rua Do Matão 1371, São Paulo, SP, 05508-090, Brazil}

\emailAdd{caina.oliveira@usp.br}
\emailAdd{emoura@if.usp.br}

\abstract{

As the nearest radio galaxy, Centaurus A (Cen~A) has been observed across multiple wavelengths and over a broad range of angular scales, providing valuable insights into the functioning of radio galaxies. However, despite being a known source of $\gamma$ rays, the physical mechanisms responsible for these high energy emissions remain unclear. In this study, we explore the capabilities of the Cherenkov Telescope Array Observatory (CTAO) to provide insights into the origin of the $\gamma$-ray emission from \cena. Using simulated observations based on the CTAO Instrument Response Functions (IRF) for the Alpha Configuration of telescopes, we investigate both the spectral and morphological domains to assess the observatory’s potential to constrain competing models of particle acceleration in the kiloparsec-scale jet and the Southern Giant Lobe. For the jet, we explore CTAO observations with $t_{\rm obs}=50$~h and $200$~h for different leptonic and hadronic origin scenarios for the very-high-energy $\gamma$-ray emission. We show that longer observation campaigns ($t_{\rm obs}=200$~h) would probe the presence of hard spectral components extending to tens of TeV. Morphological simulations demonstrate that CTAO can resolve the extended $\gamma$-ray emission along the jet, discriminating between point-like and Gaussian spatial models within $t_{\rm obs}=50$~h, assuming a known background model and an optimistic zenith angle $\vartheta = 20^\circ$. Under similar hypotheses, we studied the Giant Lobes evaluating both diffuse and filamentary emission templates for the S1 region detected by \textit{Fermi}-LAT. We found that CTAO-South can be sensitive to hadronic scenarios with proton energies extending beyond $10$~TeV, for the Alpha Configuration and $\vartheta=20^\circ$. Overall, our results indicate that CTAO observations of \cena can provide a decisive step toward unveiling the mechanisms driving the high-energy emission and particle acceleration in radio galaxies.

}

\begin{document}
\maketitle
\flushbottom

\section{Introduction}
\label{sec:intro}

Astrophysical environments produced by jetted Active Galactic Nuclei (AGNs) are among the most extreme in the Universe. The supermassive black hole~\cite[e.g.,][]{Ptitsyna_2016,Coimbra_Araujo_2021}, its accretion disk~\cite[e.g.,][]{Kimura_2015,Kheirandish_2021}, relativistic jet and surrounding medium~\cite[e.g.,][]{Rieger_2006,matthews_2018,Mbarek_2019,shukla2020gamma,zech_lemoine_2021,Medina_Torrejon_2021,Cerutti_Giacinti_2023} lead to a plethora of non-thermal acceleration mechanisms. The resulting non-thermal electrons produce radiation detected across the entire electromagnetic spectrum~\cite[e.g.,][]{review_agn,bose2022galactic}. Theoretically, the hadronic component can reach the highest energies observed in nature, potentially giving rise to ultra-high-energy cosmic rays (UHECRs)~\cite[e.g.,][]{matthews_jets,rieger_2022}.

Owing to advances in $\gamma$-ray astronomy provided by satellites, such as INTEGRAL~\cite{malizia_2023} and \textit{Fermi}-LAT~\cite{Abdollahi_2020}, ground-based Imaging Atmospheric Cherenkov Telescopes (IACTs), including H.E.S.S.~\cite{HESS_survey}, MAGIC~\cite{manganaro2024}, and VERITAS~\cite{Patel:2021MR}, and Water Cherenkov Detectors, such as HAWC~\cite{UrenaMena:2023PN} and LHAASO~\cite{Wang_2024}, jetted AGNs are now well-established sources of TeV $\gamma$-rays. Nevertheless, much remains unknown about the origin of the very-high-energy (VHE) $\gamma$-radiation. The nature of the emitting particles (leptonic or hadronic), the specific region in which the acceleration occurs, and the acceleration mechanisms involved (e.g., shock, shear, magnetic reconnection) remain subjects of ongoing debate~\cite[e.g.,][for reviews]{review_agn,matthews_jets,bose2022galactic}.

AGNs are detected from radio to VHE $\gamma$-rays. The use of multi-wavelength data has been shown to be a valuable tool in extracting information about these sources. However, theoretical models describing emission in the most energetic regimes remain poorly constrained~\cite{rieger_levinson_2018,radiogal_tev_review, bose2022galactic}. Combining high-quality $\gamma$-ray observations with simultaneous or contemporaneous multi-wavelength data is essential to better constrain both the physical nature and the precise location of the $\gamma$-ray emission~\cite{cta_ksp}.

The Cherenkov Telescope Array Observatory (CTAO) is a next-generation IACT currently under construction. Owing to its enhanced angular and energy resolutions, wider energy coverage, and greater sensitivity, the CTAO is expected to significantly advance our knowledge of extreme astrophysical environments. These may be associated with the origin of highly energetic cosmic particles -- a possible link, which will be studied as part of the CTAO \textsc{Key Science Projects}~\cite{cta_ksp}.

Radio galaxies constitute a distinct subclass of jetted AGNs, whose jet is not aligned with the observer's line of sight. The lateral view allows for the study of the structures along the jet and of the surrounding environment across multiple wavelengths. A suitable object to test the potential of the CTAO in probing particle acceleration in extragalactic environments is the radio galaxy Centaurus~A (Cen~A). \cena is the nearest radio galaxy in our cosmic neighborhood, located at a distance of $3.8 \pm 0.1$~Mpc~\cite{harris2010distance}. Its energy spectrum and morphology have been extensively explored by several instruments across multiple wavelengths, including recent observations by the Event Horizon Telescope (EHT)~\cite{janssen2021event}, Murchison
Widefield Array (MWA)~\cite{mckinley2022multi}, \textit{Swift}/BAT~\cite{Rani_2022}, \textit{Chandra}~\cite{Bogensberger_2024}, \textit{Fermi}-LAT~\cite{hess_fermi_core}, H.E.S.S.~\cite{fermi2016,hess_fermi_core,hess2020resolving}, and multi-wavelength studies~\cite[e.g.,][]{Goodger_2010,hess_fermi_core,borkar_2020,Ehlert_2022}. \cena is also at the center of the debate regarding the origin of UHECRs, as it has been proposed as a potential contributor to the excess of events observed in the so-called \textit{Centaurus region} or even the full UHECR spectrum~\cite[e.g.,][]{ROMERO1996279,Mollerach_cenA,deoliveira2025_history}.

Since the late 1990s~\cite{Hartman_1999}, \cena has been known as a source of high-energy radiation. Up to a few GeV, the observed emission is well described by a simple single-zone synchrotron-self Compton (SSC) model~\cite{hess_fermi_core}. Precise measurements of \textit{Fermi}-LAT and H.E.S.S.~\cite{hess_fermi_core} revealed that the energy spectrum hardens above a few GeV, without significant variability, indicating the presence of an additional $\gamma$-ray emitting component. Although this excess was initially interpreted as arising from an additional core or jet component, recent results of H.E.S.S.~\cite{hess2020resolving} have revealed spatially extended $\gamma$-ray emission, suggesting particle acceleration occurring along the kiloparsec-scale jet (kpc jet).

While the jet of Cen~A is a well-established TeV emitter, $\gamma$ rays from its Giant Lobes have been detected in the range from $60$~MeV up to $30$~GeV by \textit{Fermi}-LAT~\cite{fermi2016}. The Giant Lobes of \cena span approximately $8^\circ$ in radio observations, corresponding to a projected linear size of $\sim480$~kpc~\cite{mckinley2022multi}. Analysis of seven years of \textit{Fermi}-LAT data has shown that both the spectral index and the energy flux vary between the Northern and Southern lobes, as well as with the distance from the galactic nucleus.

Despite the observations, the origin of the $\gamma$ rays remains unclear, both for the kpc jet and for the Giant Lobes. Given the scientific importance of Cen~A, a deep-exposure observation of it with CTAO is planned~\cite{cta_ksp}. This program is dedicated to observing both the central kpc jet and the giant radio lobes. Regarding the central source, the main goal is to distinguish between the different interpretations for the $\gamma$-ray excess detected by \textit{Fermi}-LAT and H.E.S.S. For the giant lobes, the objective is to search for diffuse TeV emission. To cover the core and the Southern radio lobe, at least three pointings have been proposed, each with an exposure of approximately $50$~h, employing all CTAO telescopes and resulting in a total observing time of about $150$~h. The program may be extended to include observations of the Northern lobe if a diffuse emission is detected in the Southern one.

%========================

In this study, we explore the capability of the CTAO to improve our understanding of the $\gamma$-ray signals previously detected by \textit{Fermi}-LAT and H.E.S.S..
Section~\ref{sec:jet} focuses on the $\gamma$-ray signal of the kpc jet. We investigate the emission expected to be detected by the CTAO for different observation times and zenith angles. Particular attention is given to the capability of the CTAO to identify the origin of the $\gamma$-ray radiation by testing the predicted signal for several proposed scenarios of very-high-energy (VHE) emission. In Section~\ref{sec:lobes}, we examine the potential $\gamma$-ray signal from the Southern Giant Lobe based on extrapolations of the \textit{Fermi}-LAT flux, considering different maximum particle energies and emission morphologies. The main conclusions of this work are summarized in Section~\ref{sec:conclusion}.

\section{Perspectives for the central emission} \label{sec:jet}
As discussed above, the kpc jet of \cena is a well-established TeV $\gamma$-ray emitter. Based on previous measurements by \textit{Fermi}-LAT and H.E.S.S., we assessed the capability of the CTAO to detect the $\gamma$-ray signal from the kpc jet of \cena for different observation times and zenith angles. Initially, \cena is treated as a point source. The implications of this assumption are discussed further in Section~\ref{sec:3d:H.E.S.S.}, where we show that accounting for the spatial extension of the source has only a minor effect on the determination of the kpc-jet energy spectrum. A model for the spectral energy distribution (SED) in the TeV range was developed by the H.E.S.S. Collaboration~\cite{hess2020resolving} by fitting the synchrotron and Inverse Compton (IC) emission of a population of electrons along the jet. The electrons follow an exponential-cutoff broken power-law energy distribution. The SED of the kpc-scale jet is calculated considering synchrotron and inverse-Compton emission, with the jet synchrotron emission, the extended dust disk, galactic starlight, and the Cosmic Microwave Background providing seed photons for IC scattering. This model is adopted as the fiducial description of the VHE $\gamma$-ray flux in the present study and is hereafter referred to as the \textit{H.E.S.S. model}.

Simulated observations of the energy spectrum based on the \textit{H.E.S.S. model} were performed using the publicly available package \texttt{gammapy 2.0}~\cite{gammapy:2023,gammapy:zenodo-2.0}. We employed the Instrument Response Functions (IRFs) for the CTAO corresponding to the Alpha Configuration~\cite{ctao_prod5v01}. The Alpha Configuration assumes an array layout of 14 Medium-Sized Telescopes (MSTs) and 37 Small-Sized Telescopes (SSTs) at the southern site (CTAO-South); and 4 Large-Sized Telescopes (LSTs) together with 9 MSTs in the northern site (CTAO-North). Performance estimations are available for zenithal angles $\vartheta = 20^\circ,~40^\circ,~\rm{and}~60^\circ$.~\footnote{Considering the Azimuth-averaged IRFs.}

\begin{figure}[htbp]
\centering
\includegraphics[width=.8\textwidth]{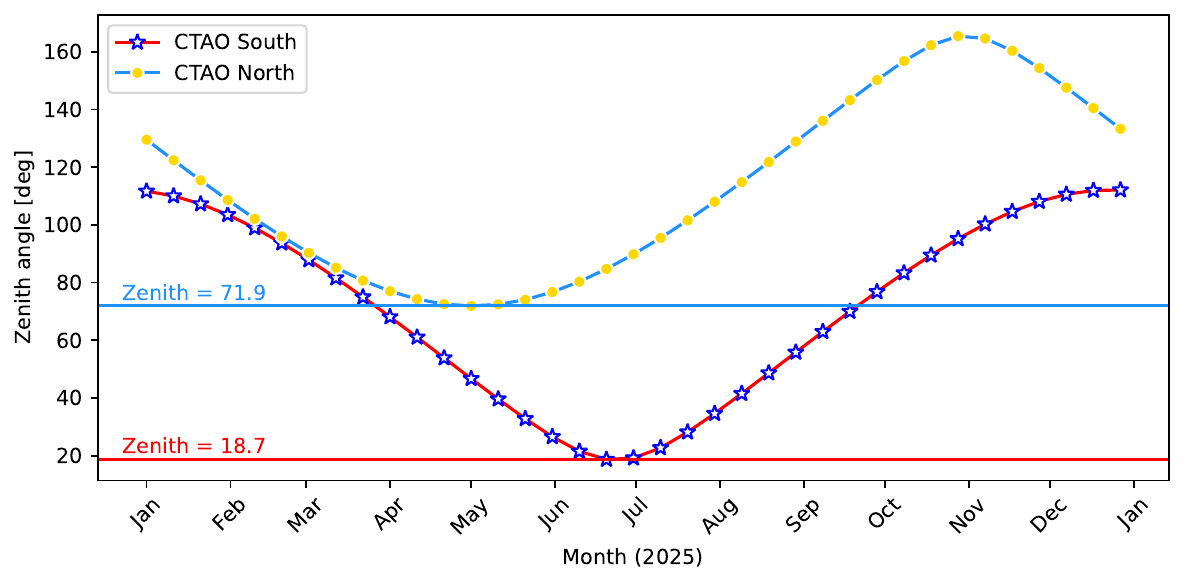}
\caption{Zenith angle of \cena throughout the year for the two CTAO's sites. The minimum zenith for each site is indicated by a horizontal line. The year 2025 is used as a representative example for the analysis.}
\label{fig:zenith}
\end{figure}

We began by exploring the appropriate IRFs to simulate observations with the CTAO-North and CTAO-South sites. Figure~\ref{fig:zenith} shows the zenith angle of \cena throughout the year for both CTAO locations. As expected from its declination, the source is best observed from the southern site, while it remains barely above the horizon for the northern site, with $\vartheta \gtrsim 70^\circ$. For CTAO-South, \cena is observed for $\sim6$~months for $\vartheta \lesssim 60^\circ$ and $\sim1$~month for $\vartheta \sim20^\circ$. Consequently, our analysis focuses on observations from CTAO-South with $\vartheta = 20^\circ, ~40^\circ$, and $60^\circ$.

We performed an On-Off analysis, assuming that \cena can be approximated as a point source, with an on-radius of $0.11^\circ$ and a telescope offset of $0.5^\circ$ with respect to the source's true position. This is a suitable approximation, as the semi-major axis determined by H.E.S.S. is $\sigma = 0.041^\circ$, which is significantly smaller than the selected on-region. The acceptance of the off-region was set to $10$ times that of the on-position. Flux points were derived using the \texttt{FluxPointsEstimator} function, with statistical errors and upper limits computed at $1\sigma$ and $2\sigma$ levels, respectively. The flux estimation procedure was repeated $1~\!000$ times. For each realization, the test statistic (TS) is obtained considering the \texttt{WStat} statistic in \texttt{gammapy}, defined as $\rm{TS} = -2 \ln \left( \mathcal{L}_{\rm max}({\rm H_0}) / \mathcal{L}_{\rm max}({\rm H_1}) \right)$. Here, TS represents the log-likelihood ratio evaluating whether the data are better described by the background-only model (null hypothesis, $\rm{H_0}$) or by the addition of a source component (alternative hypothesis, $\rm{H_1}$). Throughout this work, the reported $\rm{TS}$ corresponds to the stacked value summed across all energy bins.

\begin{figure}[htbp]
\centering
\includegraphics[width=1\textwidth]{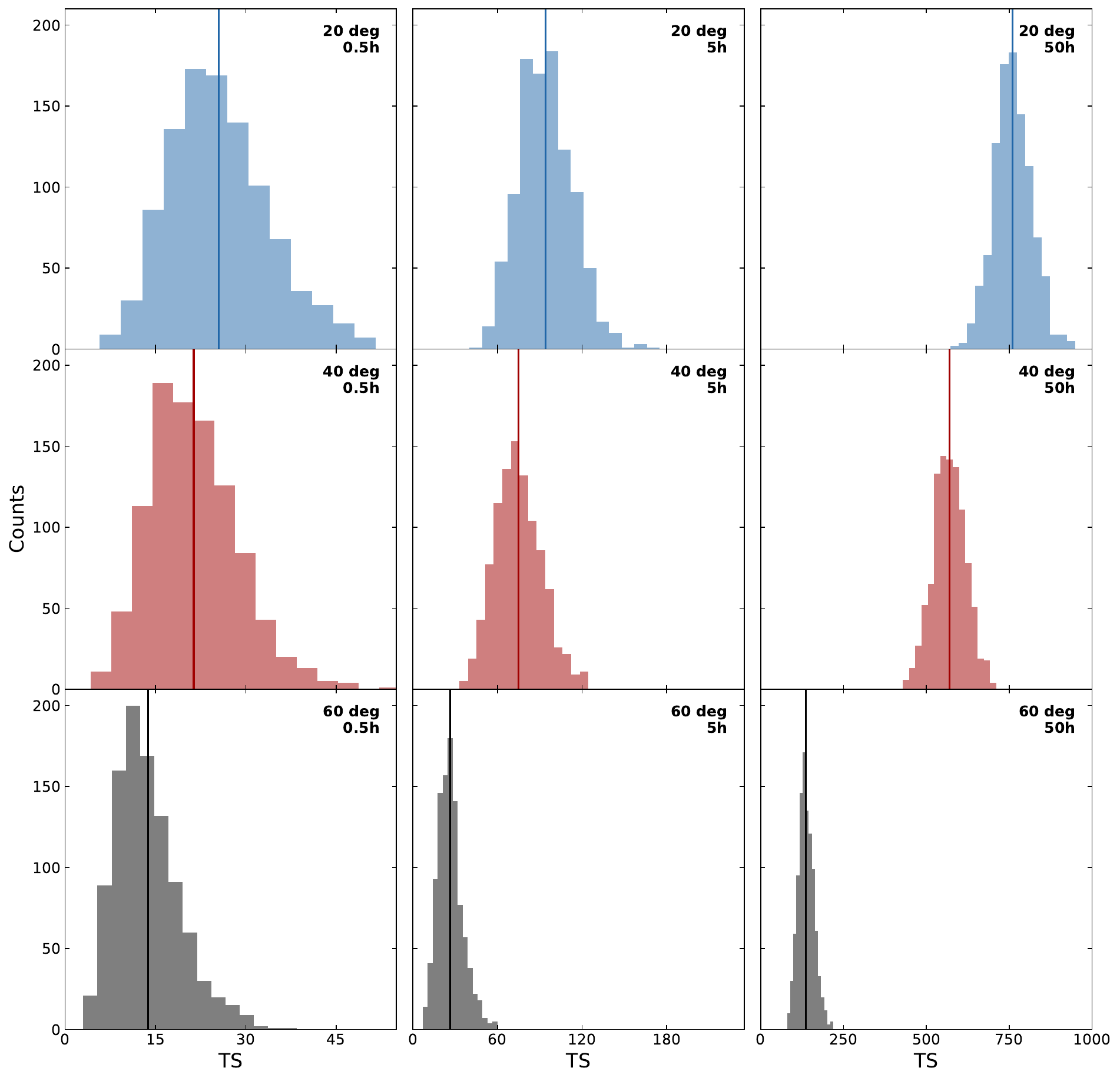}
\caption{TS distributions for $1000$ simulations of \cena as observed by the CTAO, shown as a function of live time and zenith angle. The kpc jet of \cena is approximated by a point source. Vertical bars indicate the observation whose TS values are closest to the mean value (\textit{typical observation}).}
\label{fig:hist_TS}
\end{figure}

\begin{figure}[htbp]
\centering
\includegraphics[width=0.5\textwidth]{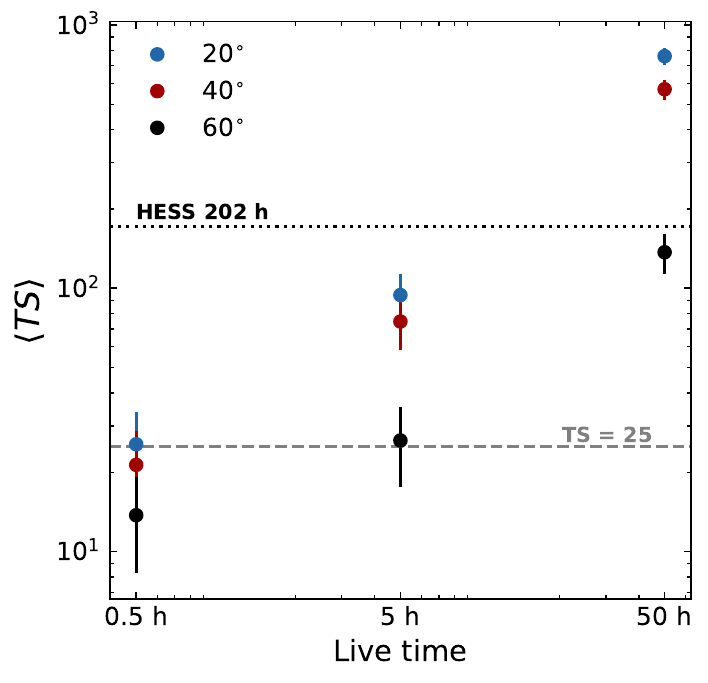}
\caption{Mean TS calculated from $1000$ simulations of \cena as observed by the CTAO as a function of live time. Error bars indicate one standard deviation between realizations. Blue, red, and black stand to $\vartheta = 20^\circ, ~40^\circ, ~ \rm{and} ~60^\circ$, respectively. The gray dashed line indicates $TS = 25$ ($5\sigma$ detection), while the black dotted line marks a $13.1\sigma$ detection as reported by the H.E.S.S. Collaboration~\cite{hess2020resolving}.}
\label{fig:mean_TS}
\end{figure}

Figure~\ref{fig:hist_TS} shows the TS distribution predicted for CTAO observations of \cena at zenith angles $\vartheta = 20^\circ, ~40^\circ, ~ \rm{and} ~60^\circ$ and live times of $t_{\rm obs} = 0.5, ~5, ~\rm{and} ~50$~h.\footnote{Along this work, we refer to $t_{\rm obs}$ as the total observation time, independent from the number of appointments needed to achieve it.} Comparing histograms at fixed zenith angle (rows) shows that the distributions shift toward higher values as the live time increases (note the different TS scale among different observation times). In contrast, increasing $\vartheta$ (columns) leads to lower TS values and produces a more gradual shift with increasing observation time. The mean TS obtained for each configuration is presented in Figure~\ref{fig:mean_TS}. An average significance level of $5\sigma$ can be achieved for $t_{\rm obs} \sim 0.5 - 5$~h, depending on the zenith angle. The current $\sim 13\sigma$ significance obtained by H.E.S.S. after $202$~h of observation~\cite{hess2020resolving} should be reached for $t_{\rm obs} \sim 5-50$~h at $\vartheta = 20^\circ - 40^\circ$.

\begin{figure}[htbp]
\centering
\includegraphics[width=1\textwidth]{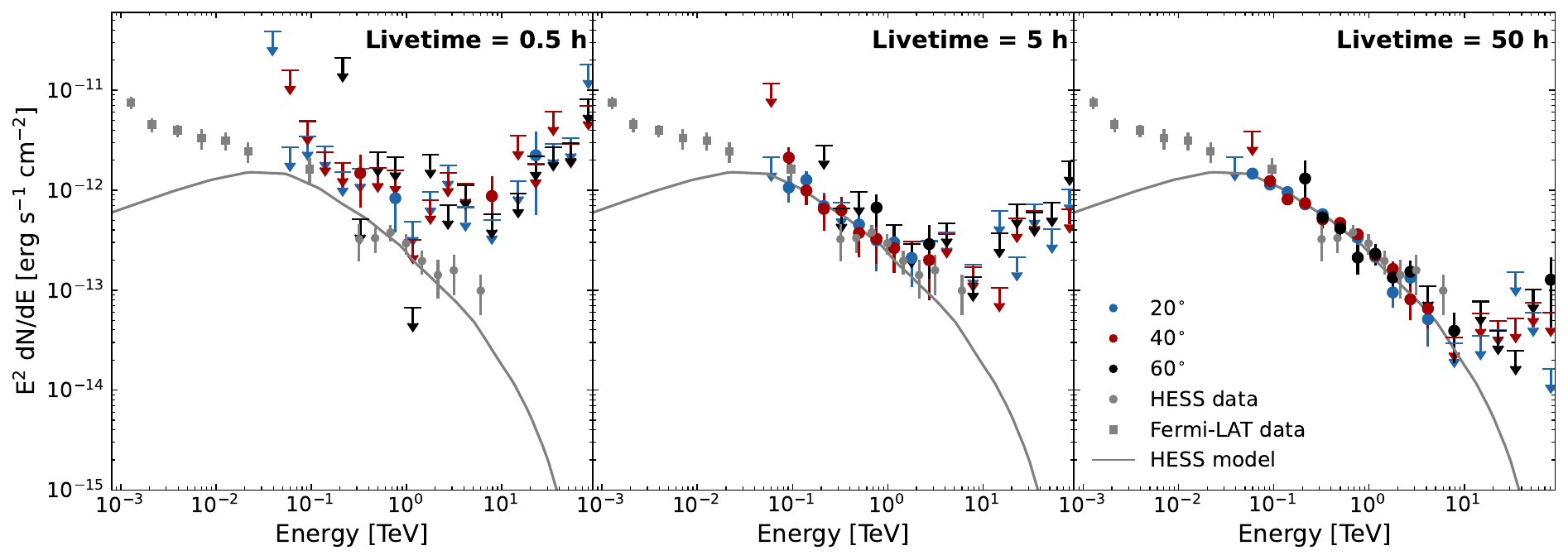}
\caption{Predicted VHE $\gamma$-ray spectra for CTAO observations based on the H.E.S.S. flux model~\cite{hess2020resolving}. Live times of $0.5,~5, \rm{~and} ~50$~h are shown in the left, center, and right panels, respectively. Blue, red, and black points stand for $\vartheta = 20^\circ, ~40^\circ, ~ \rm{and} ~60^\circ$, respectively. \textit{Fermi}-LAT and H.E.S.S. data points~\cite{hess_fermi_core} are shown as gray squares and circles, respectively. The H.E.S.S. model, used as the true spectrum to generate the CTAO synthetic data, is represented by the gray line.}
\label{fig:spec}
\end{figure}

Figure~\ref{fig:spec} presents the energy spectrum obtained for the realization with TS value closest to the mean (see Figure~\ref{fig:hist_TS}), hereafter referred to as the \textit{typical observation}. The number of energy bins was selected to match those used for the H.E.S.S. analysis~\cite{hess2020resolving}. The expected CTAO data become comparable to the H.E.S.S. measurements for $t_{\rm obs} \sim 5$~h at $\vartheta = 20^\circ - 40^\circ$. Longer observations of $t_{\rm obs} = 50$~h yield significantly improved precision, even for $\vartheta = 60^\circ$.

\subsection{Origin of the VHE radiation and more energetic channels} \label{sec:spec:probing}

Despite its detection, the origin of Cen~A's VHE $\gamma$-rays remains uncertain~\cite{hess2020resolving}, and several theoretical models have been proposed to explain the observations (e.g., \cite{rieger_levinson_2018,radiogal_tev_review,deoliveira_review} for reviews). These models vary in the physical origin of the radiation (leptonic, hadronic, or lepto-hadronic), the acceleration site (confined vs. extended), and the level of physical detail incorporated. We select several scenarios potentially consistent with the morphology and flux recently measured by H.E.S.S. to evaluate CTAO’s capability to detect and discriminate among their predicted signals. To this end, we simulate CTAO observations with fixed parameters of $t_{\rm obs} = 50$~h and $\vartheta = 20^\circ$ for three representative models of the $\gamma$-ray flux:

\begin{itemize}
    \item \textbf{H.E.S.S. model}~\cite{hess2020resolving}: the one reported by the H.E.S.S. Collaboration, in which VHE $\gamma$-rays originate from Inverse Compton scattering of electrons accelerated along the jet;
    \item \textbf{Knot -- AX1C}: shock complexes around radio/X-ray knots are proposed as efficient accelerators of electrons~\cite{Tanada_2019,Sudoh_2020,knots_2025}. We adopt the predictions for the knot AX1C (Fig. 13 of \cite{knots_2025}), where the acceleration efficiency is low ($\sim10^{-3}$), and the maximum electron energy is $\sim40$~TeV;
    \item \textbf{Jet -- B+20}~\cite{banik_cenA_2020}: lepto-hadronic scenario, where electrons, protons, and nuclei can be accelerated along the jet of Cen~A. VHE $\gamma$ rays are produced by leptonic and hadronic emission of particles accelerated in different blobs along the jet. For energies above $\sim1$~TeV, the hadronic emission dominates the flux.
\end{itemize}

In addition to the energy range already probed by \textit{Fermi}-LAT and H.E.S.S., from tens GeV up to tens TeV, the existence of components extending into the $10-100$~TeV range has been explored theoretically (e.g., \cite{radiogal_tev_review,deoliveira_review} and references therein). To assess the capability of CTAO to explore this higher-energy domain, we also include the following models in the analysis:

\begin{itemize}
    \item \textbf{Knot -- BX}: shock complexes formed by the interaction between the jet and stellar winds can accelerate electrons up to a few PeV in the case of highly efficient particle acceleration. This process would produce an accompanying $\gamma$-ray emission extending up to $10-100$~TeV~\cite{knots_2025}. We adopt the extreme scenario presented in ref.~\cite{knots_2025}, where electrons are accelerated with full efficiency ($\chi=1$) in the BX knot complex;
    \item \textbf{Jet -- Z\&M 23}: Acceleration of hadrons up to the ultra-high-energy regime along the jet can result in the emission of photons with a peak around $\sim10-100$~TeV. As proposed in ref.~\cite{zhang_murase_2023}, the resulting emission depends on the type of nuclei being accelerated. We consider the scenario where oxygen nuclei are accelerated;
    \item \textbf{Propagation -- O\&S 21}: The extragalactic propagation of UHECRs from Cen~A and the resulting electromagnetic cascades are expected to produce VHE gamma-ray radiation peaking at $\sim10–100$~TeV \cite{de2021probing}. The intensity and morphology of this flux depend on both the extragalactic magnetic field (EGMF) and the primary cosmic-ray composition. Under these conditions, the radiation is not directly associated with the jet itself, but is instead generated during propagation between the source and Earth. Consequently, the expected signal is diffuse, spanning an angular extension that depends on the EGMF strength. To evaluate the prospects of detecting this secondary photon signal with the CTAO, we compare this scenario against other models. In this study, we examine the flux predicted by the AstroR magnetic-field model \cite{hackstein2018} assuming a light, solar-like composition (C1) from ref.~\cite{de2021probing}.
\end{itemize}

At energies between $\sim1–10$~TeV, the flux from these higher-energy components drops below the H.E.S.S. measurements. To account for the emission in this lower-energy region, these fluxes were superposed onto the prediction from the \textit{H.E.S.S. model}. In particular, the combined flux from the BX model and the H.E.S.S. baseline was rescaled by an arbitrary factor of $0.75$ to ensure the total emission remained within the experimental uncertainty band of the H.E.S.S. data.

\begin{figure}[htbp]
\centering
\includegraphics[width=1\textwidth]{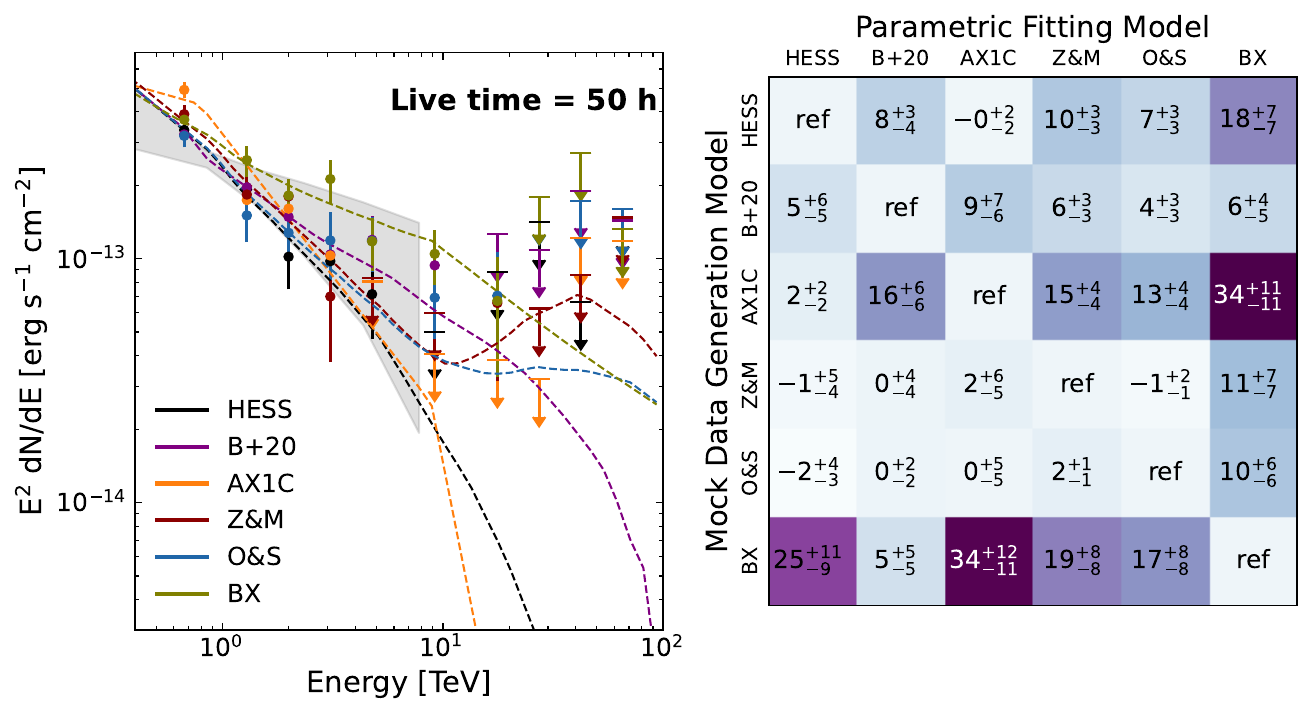}
\caption{Comparison of models of the VHE $\gamma$-ray emission for $t_{\rm obs}=50$~h. \texttt{Left}: Typical CTAO observations simulated under different assumptions for the origin of the VHE $\gamma$-rays. The gray butterfly represents the H.E.S.S. data~\cite{hess2020resolving}. Each set of flux points corresponds to a \textit{typical observation} generated using the spectral model of the same color (dashed lines).
\texttt{Right}: Mean and one standard deviation of $\chi^2 - \chi^2_0$ values obtained from $1000$ realizations of the flux predicted for CTAO observations of different models in the high-energy regime. Each row represents the data generated from one model. Each column shows the resulting fit of the data using alternative models.}
\label{fig:conf_lower}
\end{figure}

For each model, $1~\!000$ realizations of the flux estimation were performed for energies between $0.5$ and $100$~TeV. The resulting \textit{typical observations} and their corresponding spectral models are depicted in Figure~\ref{fig:conf_lower} (left). The SED estimated from each mock observation was then fitted using all models. The only free parameter of the fits is the flux normalization of each model, i.e., the spectral shapes are kept fixed. The values of $\chi^2 - \chi^2_{0}$ obtained from all realizations are presented in Figure~\ref{fig:conf_lower} (right) in the form of a confusion matrix. Here, $\chi^2_{0}$ refers to the model used to generate the simulated data and serves as the reference value. Rows represent the different models used during the generation process of the synthetic data, whereas columns are associated with the different parametric models fitted to the data.

We first compare the models using the fluxes in the GeV--TeV band, represented by the $3 \times 3$ upper-left block. The fluxes predicted by the \textit{H.E.S.S.} and AX1C models exhibit a high degree of similarity, yielding comparable $\chi^2$ values and implying that these scenarios cannot be easily distinguished based solely on their energy spectra. In contrast, the B+20 model predicts a higher flux at energies $\sim10$~TeV, which results in a larger $\chi^2$ value. However, the limited statistics in this high-energy regime lead to significant variance among individual realizations, challenging any clear distinction from the alternative models. Nonetheless, this elevated flux suggests that the B+20 scenario could become distinguishable for longer observation times.

Within $50$~h of observation, the data simulated using the Z\&M and O\&S models remain indistinguishable from the \textit{H.E.S.S. model}. This is because the peak predicted by these models occurs in the $\sim30–50$~TeV range, a region poorly probed by CTAO under this exposure time. However, the more stringent upper limits derived from observations of the \textit{H.E.S.S.} and AX1C scenarios can effectively bound the flux in this energy range, thereby constraining the predicted photon flux from the Z\&M and O\&S models.

Among the models evaluated for the higher-energy range, the optimistic scenario for the BX knot is the only one that can be clearly discriminated, as it predicts a distinct spectral suppression at higher energy ($\sim10$~TeV). However, this feature could still be confused with the high-energy cutoff characteristic of the hadronic B+20 model.

These results demonstrate that for observation times up to $50$~h, CTAO can be sufficiently sensitive to probe the energy range from tens of GeV up to tens of TeV in unprecedented detail. However, it remains unlikely to discriminate between different models for the VHE origin or to detect new VHE radiation components. To explore the full potential of CTAO, we extended the observation time to $200$~h. This observation time is comparable both to the total exposure currently achieved by H.E.S.S. and to the $150$~h allocated in the initial observation program proposed by the CTAO Collaboration. Consequently, a $200$~h observation represents a plausible extension of the initial observing strategy.

\begin{figure}[htbp]
\centering
\includegraphics[width=1\textwidth]{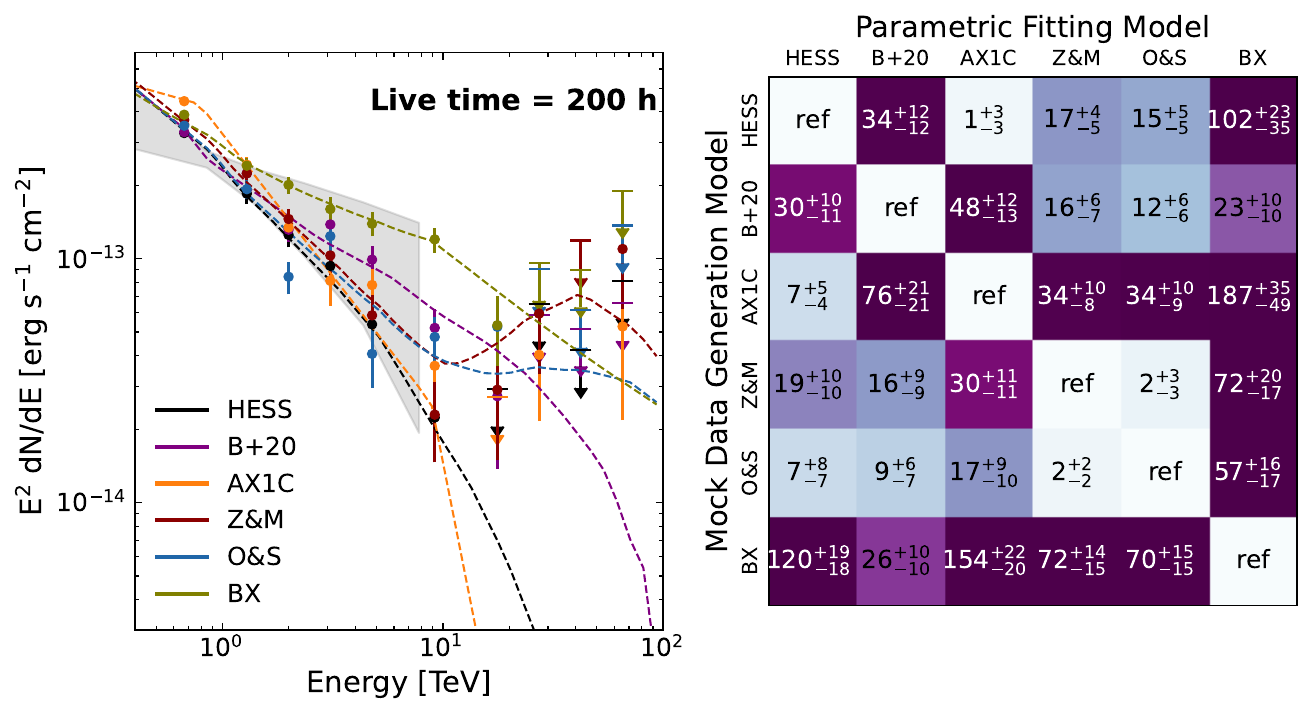}
\caption{Comparison of models of the VHE $\gamma$-ray emission for $t_{\rm obs} = 200$~h. The elements in the figure are the same as in Figure~\ref{fig:conf_lower}. }
\label{fig:conf_higher}
\end{figure}

Figure~\ref{fig:conf_higher} displays a representative simulated observation of the energy spectrum for each scenario (left) alongside its corresponding confusion matrix (right). With a longer observation time and the optimistic assumptions adopted here, it becomes possible to distinguish among the purely leptonic scenarios --- such as the \textit{H.E.S.S.}, AX1C, and BX models --- from those featuring a hadronic component (B+20, Z\&M, and O\&S). In the case of a leptonic signal based on BX model, CTAO should observe a smooth decline in the flux above $10$~TeV. Conversely, a hadronic signature would manifest as a peak in the tens-of-TeV range for Z\&M, and O\&S models. The flux from the O\&S model produces a slight spectral hardening around $10$~TeV, though detecting the emission at even higher energies remains difficult. More interestingly, it becomes possible to resolve the high-energy peak predicted by the Z\&M model. From this perspective, CTAO is clearly expected to detect a new component in the VHE $\gamma$-ray spectrum. It should also be emphasized that, in general, purely leptonic and hadronic models can exhibit spectral degeneracy, rendering them indistinguishable based on energy spectrum measurements alone.

However, large uncertainties make it challenging to distinguish between \textit{in situ} acceleration (Z\&M) and the UHECR propagation signal (O\&S) based solely on the energy spectrum. Fortunately, at these extreme energies, CTAO's exceptional angular resolution provides an additional means to distinguish between these scenarios. The $\gamma$-ray signal arising from extragalactic propagation is expected to be broadened into a wide angular halo, whereas the acceleration scenario should yield a diffuse emission more aligned with the jet. As shown in Ref.~\cite{de2021probing}, the angular distribution of photons from UHECR propagation depends strongly on the properties of the extragalactic magnetic field, a detailed discussion of which is beyond the scope of this work.

\subsection{Effects of an extended source model on the flux and sensitivity}~\label{sec:3d:H.E.S.S.}

In the previous section, we approximated the VHE $\gamma$-ray emission from the kpc-scale jet of \cena as a point source. However, the H.E.S.S. Collaboration established that the signal is actually extended along the jet and is better described by a two-dimensional Gaussian, centered at right ascension ${\rm RA}\approx 13~\rm{h}~ 25~\rm{m}~ 30.3~\rm{s}$ and declination ${\rm DEC}\approx -43^\circ~00'~15^"$, with a semi-major axis of $\sigma_{\rm maj} \approx 0.041^\circ$, ellipticity $\varepsilon \approx 0.92$, and rotation angle, measured anticlockwise from
north, $\varphi \approx 43.4^\circ$~\cite{hess2020resolving}.

\begin{figure}[htbp]
\centering
\includegraphics[width=0.5\textwidth]{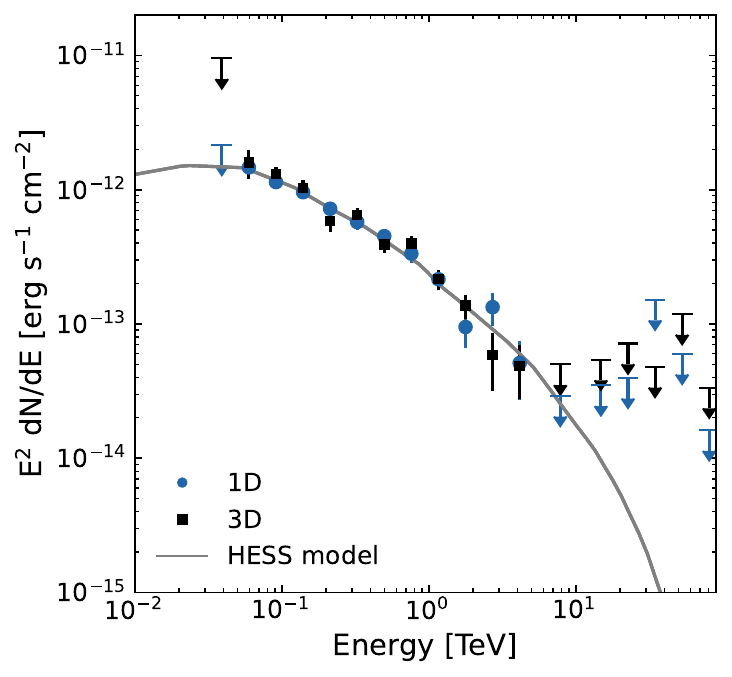}
\caption{Comparison between the typical energy spectrum of \cena considering the emission region as a point source (1D spectral analysis, blue circles) or the extended 2D Gaussian (3D spatial-spectral analysis, black squares). The gray line shows the energy spectrum model from H.E.S.S.~\cite{hess2020resolving}. Observations were simulated for $\vartheta=20^\circ$ and $t_{\rm obs} = 50$~h.}
\label{fig:3d_spec}
\end{figure}

To evaluate the impact of source extension on detectability, we simulate observations of \cena for $\vartheta = 20^\circ$ and $t_{\rm obs} = 50$~h. The spectral and spatial models adopted were the \textit{H.E.S.S. model} and a Gaussian distribution, respectively, with parameters derived by the H.E.S.S. Collaboration~\cite{hess2020resolving}. The observation geometry was defined as a $0.5^\circ \times 0.5^\circ$ window, divided into square bins of side $0.005^\circ$. This simulation was repeated 100 times, and the energy flux was estimated for each realization. For each realization, the background model is extracted directly from the IRF and evaluated across the region of interest. In this case, because a target-free background template is used rather than off-region observations, the model parameters and test statistics are evaluated using the \texttt{Cash} statistics in \texttt{Gammapy}. The typical simulation yielded a test statistic of $TS \sim 550$, which is slightly smaller than the typical values for the point-source case ($\sim750$, see Figure~\ref{fig:hist_TS}). Figure~\ref{fig:3d_spec} compares the energy spectrum of the typical realization for the point-like (Figure~\ref{fig:spec}) and extended source cases. As expected, no significant differences were found.

\begin{figure}[htbp]
\centering
\includegraphics[width=0.5\textwidth]{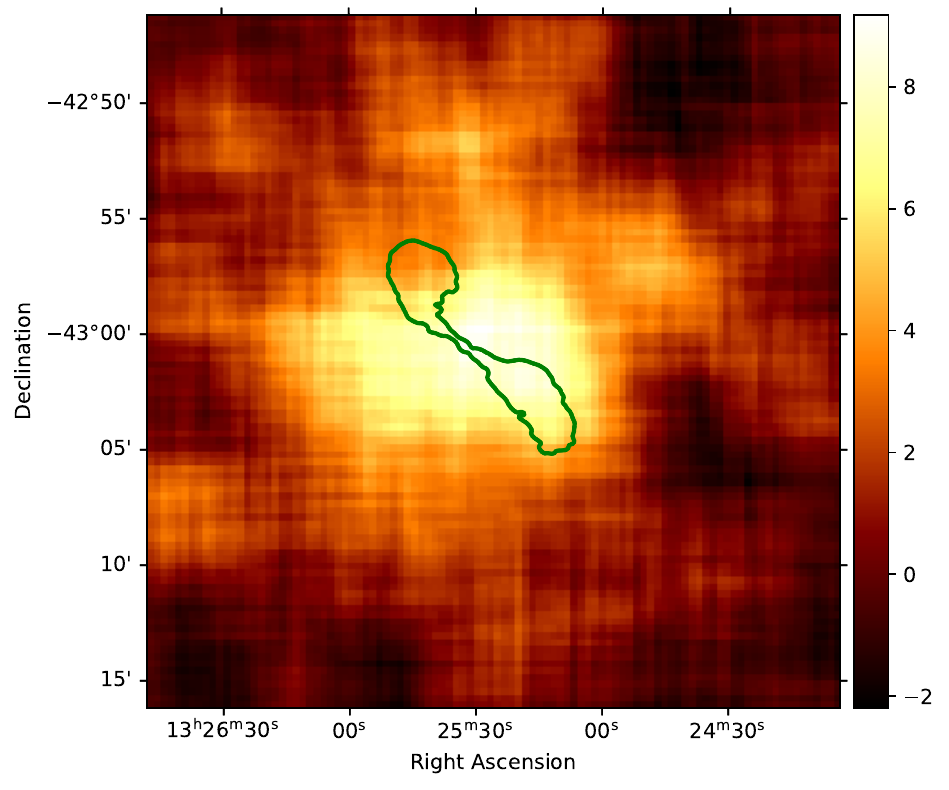}
\caption{Significance map ($\sqrt{TS}$) for the \textit{typical observation} ($t_{\rm obs} = 50$~h) of the jet of \cena, using the spatial and spectral models determined by H.E.S.S.~\cite{hess2020resolving}. A Gaussian kernel with a width of $0.08^\circ$ was applied. The green contour represents the VLA 21~cm map~\cite{condon1996}, highlighting the jet and inner lobes.}
\label{fig:H.E.S.S.:significance}
\end{figure}

\begin{figure}[htbp]
\centering
\includegraphics[width=0.9\textwidth]{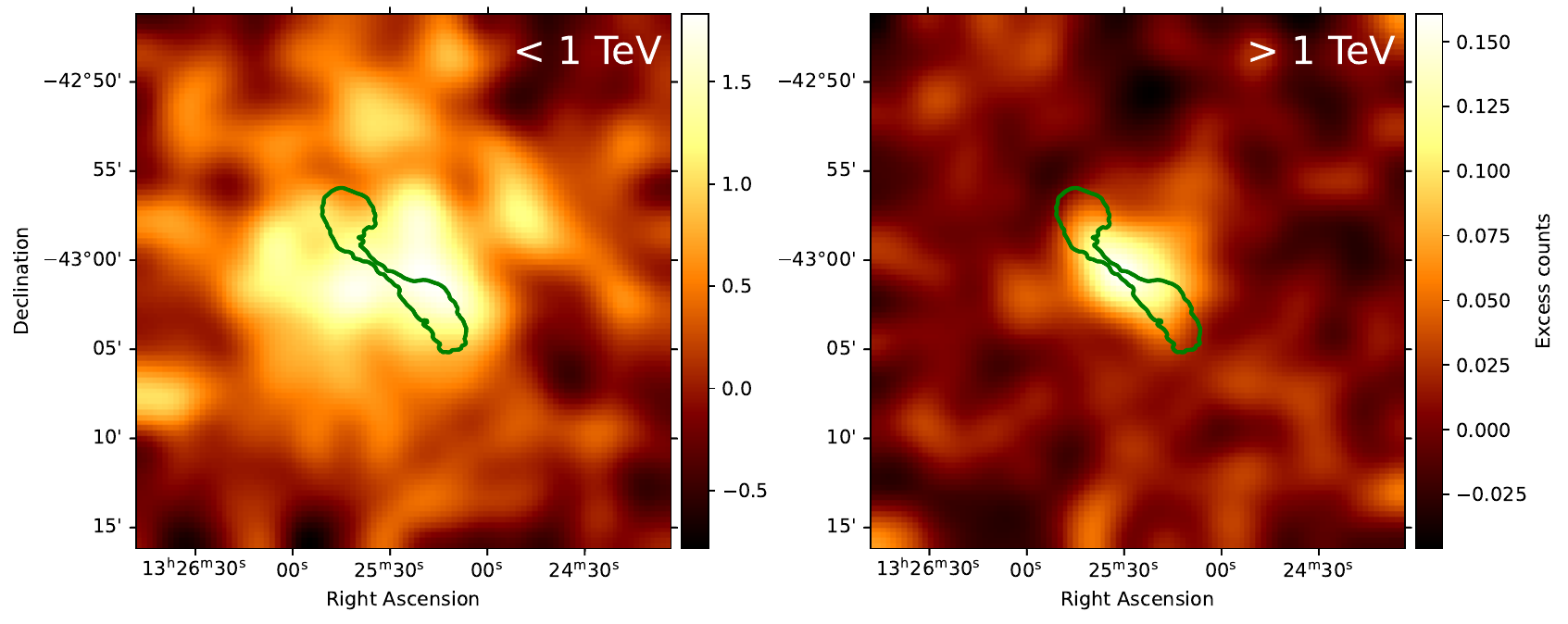}
\caption{Excess map for an observation of the inner region of \cena ($\vartheta=20^\circ$ and $t_{\rm obs} = 50$~h) for energies below (left) and above (right) $1$~TeV. The map was convolved with a Gaussian with a kernel of $0.02^\circ$. The green contour represents the VLA 21~cm map~\cite{condon1996}, highlighting the jet and inner lobes.}
\label{fig:excess}
\end{figure}

Figure~\ref{fig:H.E.S.S.:significance} shows the significance map ($\sqrt{TS}$) of a \textit{typical observation} with $t_{\rm obs}=50$~h where a Gaussian filter with $\sigma=0.08^\circ$ was applied. The map presents a peak with $\sqrt{TS}=9.1$. Figure~\ref{fig:excess} presents the corresponding excess map. Since the angular resolution of CTAO Alpha Configuration improves significantly with energy, reaching $\lesssim 0.05^\circ$ above $\sim 1$~TeV, the map is depicted for two energy bins (below and above 1~TeV). As expected, the source morphology would become visible at higher energies for $\sim50$~h observations.

\begin{figure}[htbp]
\centering
\includegraphics[width=0.5\textwidth]{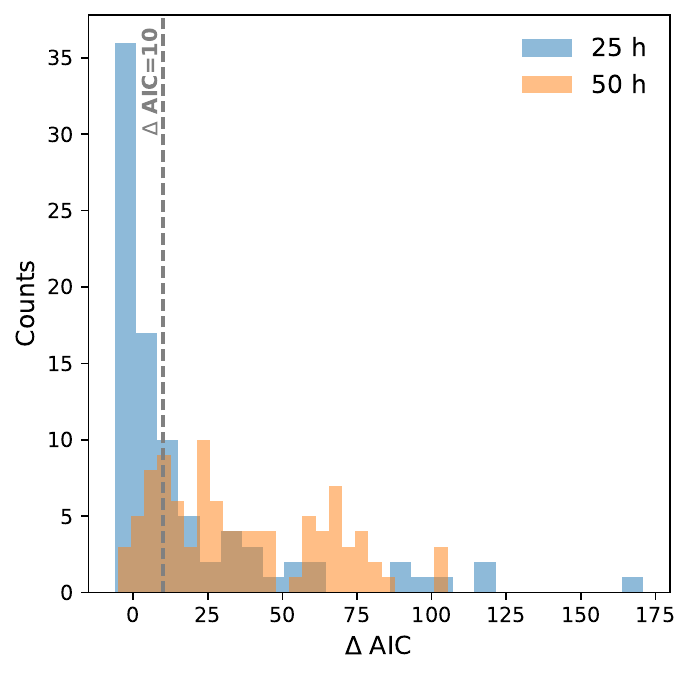}
\caption{Distribution of $\rm \Delta AIC= AIC_{point} - AIC_{gauss}$ comparing the fit of a 2D Gaussian and a point source for simulations of extense emission in the kpc-scale jet of Cen~A. Two observation times are considered, 25~h (orange) and 50~h (blue).}
\label{fig:AIC_jet}
\end{figure}

To assess whether CTAO can distinguish between a point source and a Gaussian model for the emission region, we fitted the simulated maps. The Gaussian model includes 5 free parameters (RA, DEC, $\sigma$, $\varepsilon$, $\varphi$), while the point source has 2 free parameters (RA, DEC). Applying the Akaike Information Criterion~\cite{burnham2002model}, we obtained the distribution of $\rm \Delta AIC= AIC_{point} - AIC_{gauss}$, which is shown in Figure~\ref{fig:AIC_jet}. We obtain $\rm \Delta AIC_{50 \rm h} = 36 \pm 28$, of which $76\%$ of the realizations yield $\rm \Delta AIC_{50 \rm h} > 10$, strongly favoring the Gaussian model over the point source.

To assess whether the morphology can be constrained with shorter observation times, the analysis was repeated for $t_{\rm obs} = 25$~h. It resulted in $\rm \Delta AIC_{25 \rm h} = 17 \pm 33$, with $44\%$ of the realizations with $\rm \Delta AIC_{50 \rm h} < 5$ (see Figure~\ref{fig:AIC_jet}). This suggests that observations longer than $25$~h are necessary to discriminate between the models.

\section{Perspectives for the Giant Lobes}~\label{sec:lobes}

Gamma-ray emission in the energy range $60$~MeV to $30$~GeV has been observed from the Giant Lobes of \cena by \textit{Fermi}-LAT~\cite{fermi2016}. In the analysis performed in ref.~\cite{fermi2016}, the lobes were sliced into six regions, N1, N2, N3, S1, S2, and S3, where N/S denote the Northern/Southern lobes, and $1-3$ indicate decreasing distance from the core, from the edges ($1$) toward the center ($3$). The core region was masked to avoid contamination. An important question that remains from this analysis is whether the $\gamma$ radiation has a leptonic or hadronic origin, since both models provide satisfactory fits to the data. In the leptonic case, $\gamma$ rays originate from Inverse Compton scattering of electrons on relic photons of the Cosmic Microwave Background. In the lepto-hadronic scenario, the emission is dominated by the decay of pions resulting from proton-proton interactions.

A particularly interesting and peculiar region is S1, where the radio and $\gamma$-ray components share the same spectral index, and no break or cutoff is observed in the $\gamma$-ray spectrum. While all other regions can be consistently explained by a purely leptonic model, S1 requires a magnetic field strength of $\sim13~\mu$G, about ten times higher than that inferred for the other regions. On the other hand, the lepto-hadronic interpretation demands a significantly higher energy budget in relativistic protons. In this section, we investigate how the CTAO could help clarify the nature of the $\gamma$-ray emission detected by \textit{Fermi}-LAT. We focus on evaluating the detectability of the signal assuming a hadronic origin, as this scenario predicts emission extending to higher energies, while the leptonic flux is expected to be strongly suppressed in the VHE regime.

Data collected by \textit{Fermi}-LAT do not allow the determination of a spectral break or cutoff in the proton spectrum. In Ref.~\cite{fermi2016}, it was modeled as a simple power law with a spectral index $p=2.05 \pm 0.12$ and a total energy $W_{\rm p} = 1.4^{+ 0.7}_{- 0.4} \times 10^{60}$~erg, assuming an ambient gas density\footnote{The exact value of $N_H$ is still under debate~\cite{wykes_2013,mckinley2022multi}. However, this uncertainty should affect the amplitude of the non-thermal proton spectrum and, consequently, the value of $W_{\rm p}$.} $N_H=10^{-4}~{\rm cm}^{-3}$. \textit{Suzaku} X-ray observations in the $0.5-10$~keV range potentially indicate that electrons may be accelerated up to $\gtrsim10$~TeV in compact regions of the Southern Giant Lobe, despite thermal emission cannot be ruled out~\cite{Stawarz_2013}. Motivated by this result, we assume that protons can likewise be accelerated up to energies of $\gtrsim 10$~TeV. Using the \texttt{Naima}~\cite{naima} package within \texttt{gammapy}, we reproduced the $\gamma$-ray spectral energy distribution and introduced exponential cutoffs in the proton spectrum at $E_{\rm cut}^{\rm p}=10 {\rm~ and ~}100~\rm{TeV}$. The emission detected by \textit{Fermi}-LAT can hardly be modeled as a point-like source. Based on Figure~3 of Ref.~\cite{fermi2016}, we estimate that the S1 emission is dominated by a region with a diameter of approximately $1^\circ$, centered at $\rm RA \approx 200.5^\circ,~dec \approx -45.5^\circ$. An isotropic Gaussian with $\sigma=0.5^\circ$ is adopted as the fiducial model for the S1 emission. We highlighted that small variations in the estimated direction of emission should not change the results due to the directional focus of the telescopes.

As in Section~\ref{sec:3d:H.E.S.S.}, we assume the background model provided by the CTAO IRFs as a known element in the model. An angular window of $2^\circ \times 2^\circ$ with bin size $0.02^\circ$ was adopted, and the simulations were repeated 100 times. The analysis was performed for $t_{\rm obs} = 5$ and $50$~h, covering energies between $30$~GeV and $10$~TeV, with a fixed zenith angle $\vartheta = 20^\circ$. The chosen field of view is well within that of CTAO (for comparison, the field of view of the LST, MST, and SST telescopes is $>4.5^\circ$, $>7^\circ$, and $>8^\circ$, respectively~\cite{cta_ksp}).

As discussed by Ref.~\cite{fermi2016}, the lepto-hadronic model demands a high total energy in relativistic protons. This requirement could, in principle, be alleviated if the $\gamma$-ray emission originates from filamentary structures formed from the collapse of thermally unstable plasma, where cosmic rays and magnetic fields accumulate~\cite{fermi2016}. Filamentary structures are actually known from radio measurements in the giant lobes~\cite{wykes_2014}. In addition, X-ray measurements potentially imply a highly inhomogeneous and non-uniform structure for the lobes~\cite{Stawarz_2013}. Particularly interesting are the Vertex and the Vortex filaments, both localized in the Southern Lobe~\cite{wykes_2014,wykes_2015}. These filaments exhibit slightly flatter spectra than the surrounding lobe plasma, suggesting either more recent particle acceleration or regions of enhanced magnetic field~\cite{wykes_2014}. The \textit{Suzaku} X-ray observations also suggest that particle acceleration should occur in compact regions associated with the radio filaments~\cite{Stawarz_2013}.

\begin{figure}[htbp]
\centering
\includegraphics[width=0.5\textwidth]{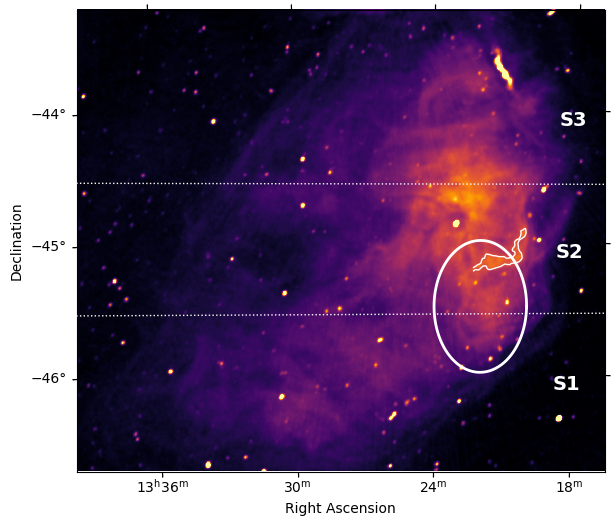}
\caption{Murchison Widefield Array radio (185 MHz) image of the Southern Lobe~\cite{mckinley2022multi}
(\url{https://cdsarc.cds.unistra.fr/viz-bin/cat/J/other/NatAs/6.109}) superposed with the characteristic morphologies of the spatial models (in white): the white circle shows the $1\sigma$ radius for the Gaussian profile, and the Vertex used as morphological template for a filamentary emission. The Vertex region was bounded by an $86\%$ contour of the maximum radio flux density in the surrounding region. Regions S1, S2, and S3 in the analysis of Ref.~\cite{fermi2016} are also shown.}
\label{fig:lobes:img}
\end{figure}

We compare the results obtained for the Gaussian model with those from a Vertex-like filament, used here as a template for filamentary emission in the lobes. Figure~\ref{fig:lobes:img} shows the representative regions of the Gaussian model with $\sigma = 0.5^\circ$ and the Vertex superposed to the radio image of the Southern Giant Lobes of Cen~A. Although the Vertex is adopted as a morphological template, we note that the Vertex itself is unlikely to be the source of the $\gamma$-ray flux detected at S1, since: (i) it is localized at the lower edge of the region S2, defined at $\rm DEC \gtrsim -45.5^\circ$, while the Vertex lies between $\rm -45.5^\circ \lesssim Dec \lesssim -45^\circ$; (ii) its origin is probably related to magnetohydrodynamic turbulence in the plasma, rather than to the collapse of thermally unstable plasma; and (iii) the X-ray flux predicted by the lepto-hadronic model for S1 greatly exceeds the upper limit of $\sim 9.6 \times 10^{-13}~{\rm erg~cm^{-2}~s^{-1}}$ at $\rm 1~keV$ measured for the Vertex.~\cite{wykes_2014,wykes_2015}

\begin{figure}[htbp]
\centering
\includegraphics[width=0.5\textwidth]{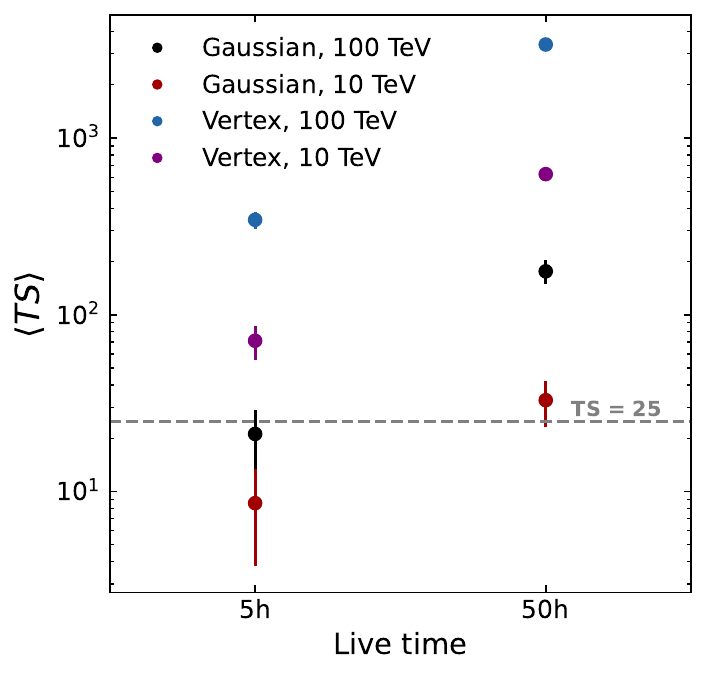}
\caption{Mean TS calculated from $100$ simulation observations of the S1 region of \cena by the CTAO as a function of live time. Error bars indicate one standard deviation across realizations. Red and black points correspond to Gaussian spatial models with $\sigma=0.5^\circ$ and spectral cutoffs of $E_{\rm cut}=10$ and $100$~TeV, respectively. Purple and blue points indicate the predicted values assuming a Vertex-like spatial model with $E_{\rm cut}=10$~TeV and $100$~TeV, respectively. In all cases, $\vartheta = 20^\circ$.}
\label{fig:lobes:TS}
\end{figure}

\begin{figure}[htbp]
\centering
\includegraphics[width=0.8\textwidth]{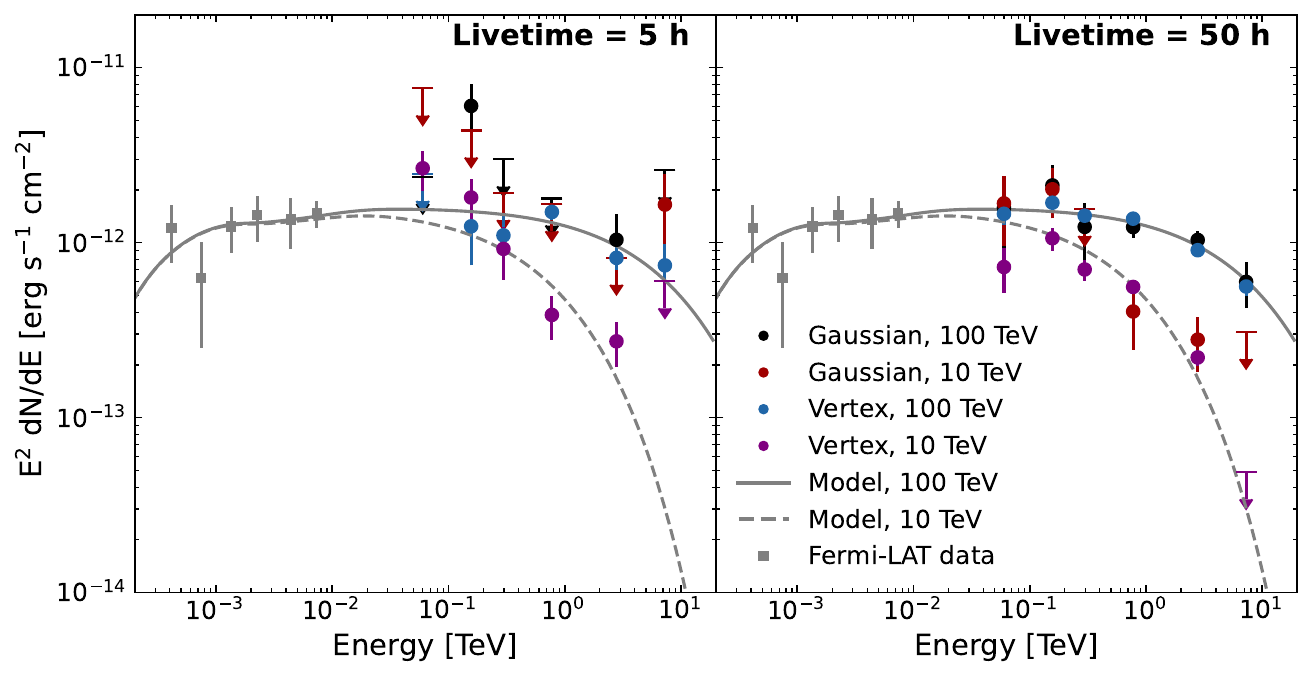}
\caption{Predicted energy spectrum of VHE $\gamma$ rays for CTAO measurements based on extrapolations of \textit{Fermi}-LAT models for region S1 of the Giant Lobes. Live times of $5, \rm{~and} ~50$~h are shown in left and right panels, respectively. Red and black points correspond to Gaussian spatial models with $\sigma=0.5^\circ$ and spectral cutoffs of $E_{\rm cut}=10$ and $100$~TeV, respectively. Purple and blue points show the predicted spectrum assuming a Vertex-like spatial model with $E_{\rm cut}=10$ and $100$~TeV, respectively. In all cases, $\vartheta = 20^\circ$. \textit{Fermi}-LAT data points are shown in gray squares. The models used as the true spectra to generate the data are shown in black ($E_{\rm cut}=10$~TeV) and red ($E_{\rm cut}=100$~TeV) lines.}
\label{fig:lobes:spec}
\end{figure}

Figure~\ref{fig:lobes:TS} shows the mean TS predicted for CTAO observations of the S1 region. The significance level is strongly dependent on both the assumed morphology (Gaussian or Vertex) and the maximum energy considered. In all cases considered, the detection exceeds $\sqrt{\rm TS} = 5$ for $t_{\rm obs} = 50$~h for the optimum angle $\vartheta=20^\circ$. Figure~\ref{fig:lobes:spec} presents the energy spectrum obtained for a \textit{typical observation} with $t_{\rm obs} = 5$ and $50$~h for $E_{\rm cut} = 10 \rm~and~100$~TeV. Since the CTAO sensitivity is optimal in the $\sim 1-10$~TeV range, both the resulting significance and fluxes are strongly affected by the assumed cutoff energy. For the Gaussian model, the spectrum is dominated by upper limits for $t_{\rm obs}=5$~h, whereas it becomes reasonably well constrained at $t_{\rm obs}=50$~h. The energy spectrum derived for the Vertex-like morphology presents less deviation from the model than that obtained with the Gaussian spatial model. This behavior is a consequence of the more localized emission produced by the filament model compared to the broader Gaussian profile. This effect is also reflected in the corresponding TS values.

To assess whether CTAO can reliably constrain the cutoff energy, we fitted the energy spectra obtained for $t_{\rm obs}=50$~h, considering the Gaussian emission profile and cutoff energies of $E_{\rm cut}=10$ and $100$~TeV. The spectra from each realisation were combined with the \textit{Fermi}-LAT flux data~\cite{fermi2016} for region S1 and fitted using a power-law model suppressed by the exponential cutoff for protons, which produces $\gamma$-rays via pion decay. The ideal case in which $s=2$ is kept fixed was considered. For $E_{\rm cut} = 100$~TeV, we obtained a reconstructed cutoff energy $E_{\rm cut}^{\rm rec} = 58_{-14}^{+16}$~TeV with a mean fit error of $19$~TeV. For the datasets generated with $E_{\rm cut} = 10$~TeV, the reconstructed values were $E_{\rm cut}^{\rm rec} = 7_{-2}^{+2}$~TeV with a mean fit error of $2.3$~TeV. These results indicate that the cutoff energy of a proton population will not be exactly reconstructed, but  can still be constrained. Similar results are obtained for a Vertex-like morphology, resulting in reconstructed values $E_{\rm cut}^{\rm rec} = 71_{-7}^{+12}$ ($8.5_{-0.7}^{+1.1}$)~TeV with a mean fit error of $11$ ($1.0$)~TeV for $E_{\rm cut} = 100$ ($10$)~TeV.

\begin{figure}[htbp]
\centering
\includegraphics[width=0.4\textwidth]{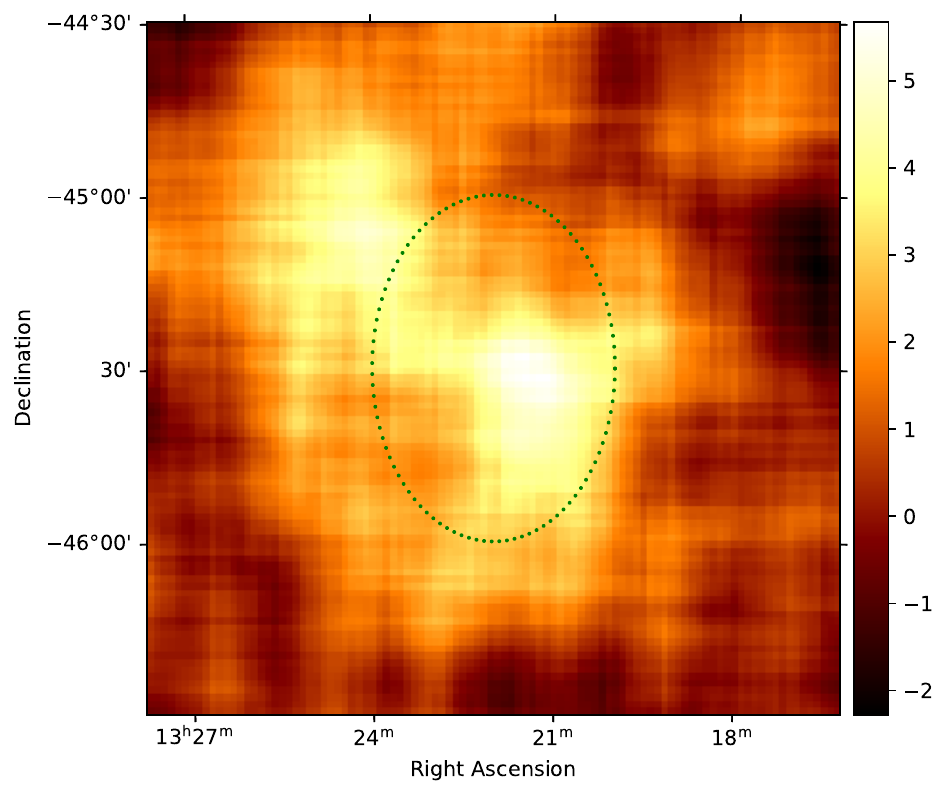}
\hspace{0.02\textwidth}
\includegraphics[width=0.4\textwidth]{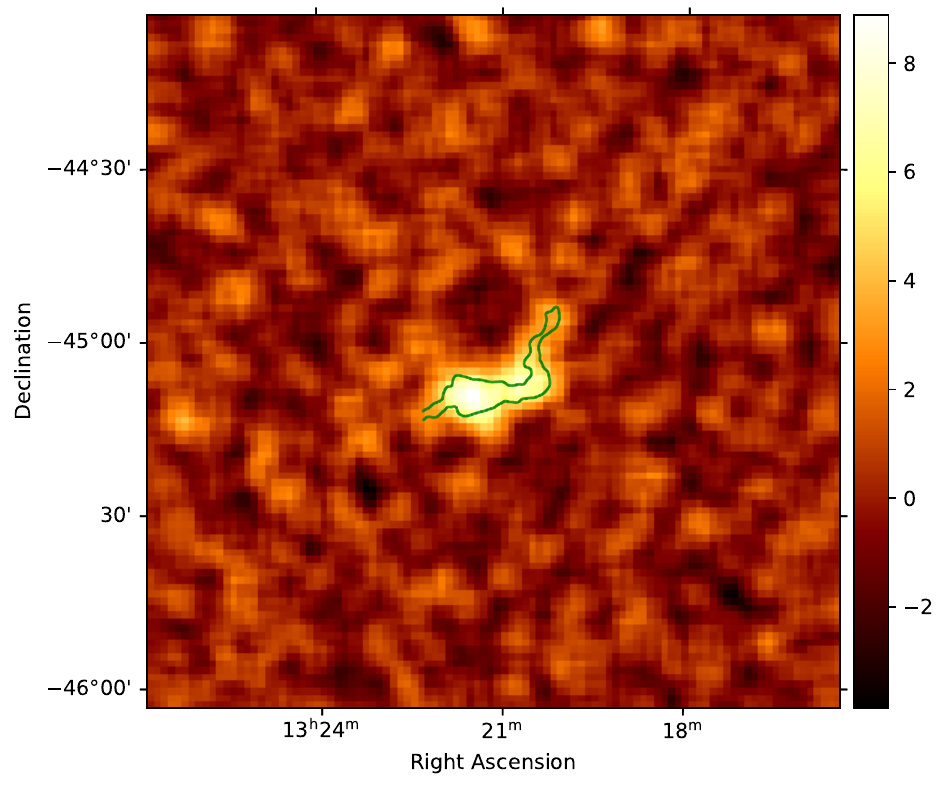}
\caption{Significance map for a \textit{typical observation} of region S1 assuming the Gaussian ($t_{\rm obs}=50$~h, left) and Vertex-like ($t_{\rm obs}=5$~h, right) morphologies ($E_{\rm cut}=100$~TeV). The significance is computed for energies $E>0.5$~TeV. The map was convolved with a Gaussian kernel of $0.5^\circ$ and $0.1^\circ$ for the Gaussian and Vertex morphologies, respectively. The green contour indicates the spatial templates.%
}
\label{fig:lobes:significance}
\end{figure}

\begin{figure}[htbp]
\centering
\includegraphics[width=0.9\textwidth]{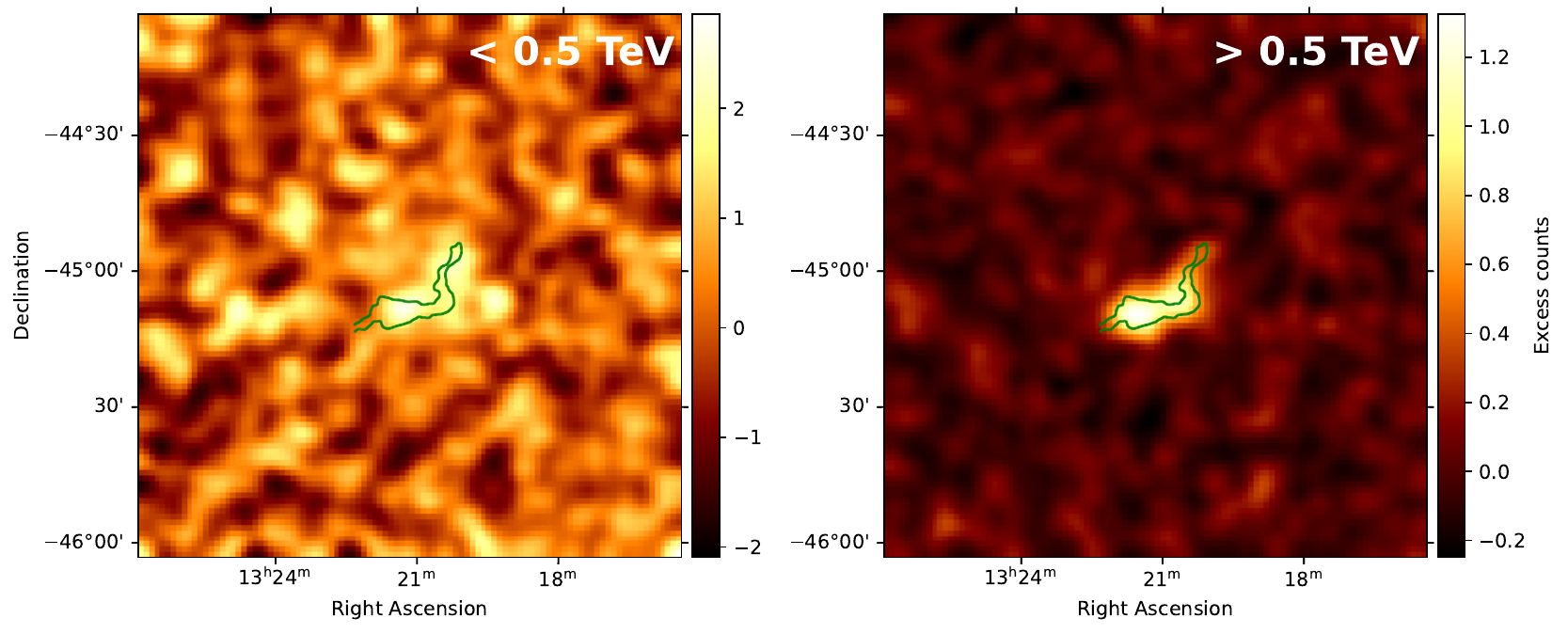}
\caption{Excess maps for a \textit{typical observation} of the S1 spectrum of \cena for energies below (left) and above (right) $0.5$~TeV, considering $t_{\rm obs}=5$~h and a Vertex-like morphology ($E_{\rm cut}=100$~TeV). Each map was convolved with a Gaussian kernel of $0.04^\circ$ for smoothness. The green contour indicates the Vertex filament used as a spatial template.}
\label{fig:lobes:excess}
\end{figure}

The significance maps obtained for energies above $0.5$~TeV for a \textit{typical observation}, assuming Gaussian and Vertex morphologies ($E_{\rm cut}=100$~TeV), are shown in Figure~\ref{fig:lobes:significance}. Note that the Gaussian map corresponds to $t_{\rm obs}=50$~h, while the Vertex map corresponds to $t_{\rm obs}=5$~h. The Gaussian morphology reaches a peak significance of $\sqrt{TS} \sim 5$ after $t_{\rm obs}=50$~h, whereas the Vertex morphology peaks at $\sqrt{TS} \sim 8$ after only $t_{\rm obs}=5$~h. These results indicate promising detection prospects but also reinforce the strong dependence on the assumed morphology. We expect that the detailed structure of the emission will become apparent with future CTAO observations. To illustrate this, Figure~\ref{fig:lobes:excess} presents the excess map derived from the Vertex-like morphology for $t_{\rm obs} = 5$~h, overlaid with the radio contour of the Vertex. The higher significance and fine structure observed for the Vertex case result from the spectral plateau produced by the spectral index of $\sim 2$ inferred from the \textit{Fermi}-LAT data and the assumed cutoff energy of $100$~TeV used in this analysis.

\section{Limitations and Conclusions} ~\label{sec:conclusion}

In this study, we assessed the potential advances that observations of \cena with CTAO can potentially bring. Considering the Alpha Configuration of CTAO and its publicly available IRFs, we predict the detection signals for several models previously proposed for the origin of the $\gamma$-radiation in the kpc-scale jet and Southern Giant Lobe of Cen~A.

The existence of a VHE signal from the central region of \cena is already known from observations of \textit{Fermi}-LAT and H.E.S.S. However, its physical origin remains an open question. The detected signal served as a guideline for the analysis of the kpc jet. The significance level achieved with $t_{\rm obs} \sim 200$~h with H.E.S.S. observations is expected to be surpassed by CTAO Alpha Configuration with only $t_{\rm obs} \sim 50$~h. An observation of the core region with an exposure time of $t_{\rm obs} = 50$~h is already planned by the CTAO Consortium~\cite{cta_ksp}. The results presented in Section~\ref{sec:spec:probing} indicate that, with this observation time, the models currently proposed in the literature can be further constrained. In particular, for models with maximum energies similar to those reported by the H.E.S.S. Collaboration~\cite{hess2020resolving}, CTAO observations might help to discriminate between hadronic and leptonic scenarios.

Similar to the strong constraints imposed by the morphological results of H.E.S.S., we expected that the morphology determined by CTAO also plays a crucial role in identifying the acceleration and emission processes of relativistic particles along the jet. Combined with the spectral information, the excellent angular resolution of CTAO will be essential to further constrain the origin of the VHE $\gamma$ rays. We demonstrate that, under the adopted assumptions, $50$~h observations is sufficient for CTAO to discriminate between point-like and extended emission scenarios.

In addition to the current measurements extending up to a few TeV, some models predict that the VHE $\gamma$-ray spectrum may reach higher energies, $\sim10-100$~TeV. We investigated the potential of CTAO observations under these scenarios for a longer exposure time ($t_{\rm obs} = 200$~h). Based on this analysis, we studied a scenario with a hypothetical new higher-energy component in the VHE $\gamma$-ray spectrum. This component could originate from highly efficient electron acceleration in knots, from the acceleration of nuclei along the jet, and/or from the propagation of UHECRs. The detection of such a signal would be used as an indirect probe of UHECR acceleration in \cena. This constitutes a remarkable and exciting prospect for high-energy astrophysics. Our results indicate a potential discrimination between the hadronic and leptonic models selected in our analysis. Nevertheless, it is important to bear in mind that, in general, purely hadronic and leptonic models are unlikely to be distinguished based on their energy spectra alone.

The Southern Giant Lobe of \cena also represents a particularly interesting site to explore the environment and particle acceleration processes occurring along the lobes of radio galaxies. The \textit{Fermi}-LAT observations raise important questions about in-situ particle acceleration and the dynamical structure of radio galaxy lobes, such as the presence of condensation filaments~\cite{fermi2016}. These questions could potentially be addressed through CTAO observations.

The absence of a cutoff in the energy spectra observed by \textit{Fermi}-LAT opens the possibility of a hadronic origin for the $\gamma$ rays in the S1 region, although it leaves the emission model within the Giant Lobes considerably more uncertain. Conversely, it also implies a potentially large energy budget in hadronic particles.  For this reason, we tested two models for the maximum energy and emission morphology in the S1 region. In all cases, we obtained favorable detection prospects, especially for longer observations ($t_{\rm obs} = 50$~h). For the current array configuration proposed for CTAO-South, the lower maximum proton energy ($E_{\rm cut} \sim 10$~TeV) begins to challenge the detectability. A filamentary, Vertex-like emission is more likely to be detected than a diffuse Gaussian with $\sigma=0.5^\circ$. We expect that even more extended emission than considered here could still be detected, provided that $E_{\rm cut}$ is not significantly smaller than $\sim10$~TeV.

The results obtained in this work can be viewed as an optimistic upper baseline for observations of Cen A with CTAO, as additional astrophysical background emission was not modeled and the zenith angle was fixed at $\vartheta = 20^\circ$. Throughout this paper, the instrumental background model is derived directly from the CTAO IRFs, consisting solely of hadronic and electronic air-shower events that depend on the field-of-view coordinates. Uncertainties in this background estimation can further degrade the detectability of the studied models. Furthermore, increasing the background contribution could impact the discrimination between point-like and extended emission, given that the variance observed for $\Delta\textrm{AIC}_\textrm{50h}$ is large. An additional complication comes from the fact that extended structures, different components of Cen A itself as well as unrelated astrophysical sources within the field of view, could act as an astrophysical background. Consequently, unmodeled field sources could further degrade the sensitivity of CTAO to detect and characterize the VHE signal from Cen~A.

The results presented in sections~\ref{sec:spec:probing}, \ref{sec:3d:H.E.S.S.}, and \ref{sec:lobes} were obtained assuming a fixed zenith angle of $\vartheta = 20^\circ$. However, realistic $50 - 200$~h observations of Cen A span a broader range of zenith angles with $\vartheta \gtrsim 20^\circ$ (see fig.~\ref{fig:zenith}). Because TS drops significantly for $\vartheta \gtrsim 40^\circ$ (fig.~\ref{fig:mean_TS}), our estimates represent an optimistic upper limit.

An additional source of variance omitted in this study is the intrinsic uncertainty within each considered model. These model uncertainties are expected to degrade the classification performance shown in figures \ref{fig:conf_lower} and \ref{fig:conf_higher}, once they exceed the statistical precision achievable with a $200$~h exposure. A special case worth noting involves the Z\&M and O\&S scenarios. Due to their relatively low flux in the $\sim 1 - 10$~TeV range, these models were combined with the H.E.S.S. baseline for comparative analysis. However, if the intrinsic flux in the $\sim 1 - 10$~TeV band is higher, such as that predicted by the BX model, it could dominate over the high-energy components of Z\&M and O\&S, rendering model discrimination substantially more challenging.

In summary, the results obtained in this work indicate, under the assumptions adopted here, that CTAO may help to make a distinction between the existing leptonic and lepto-hadronic models for the origin of the VHE $\gamma$-radiation in Cen~A. Furthermore, its unprecedented sensitivity and angular resolution can provide key morphological information on the jet and lobe structures, advancing our understanding of particle acceleration and radiation mechanisms in radio galaxies.

\acknowledgments

The authors acknowledge Vitor de Souza, James H. Matthews, M. Felipe Souza, Lívia S. Rocha, Luiz A. S. Pereira, Micael Andrade, and Julia G. Mamprim for useful discussions. We also thank Michael Zacharias, Garret Cotter, Weidong Jin, and Atreya Acharyya for carefully reading early versions of this manuscript and offering helpful comments that improved the work. This study was financed, in part, by the São Paulo Research Foundation (FAPESP), Brasil. Process Number 2025/03325-5, 2021/01089-1, and 2019/10151-2. EMS acknowledges the support of CNPq grant 311983/2022-7. The authors acknowledge the National Laboratory for Scientific Computing (LNCC/MCTI, Brazil) for providing HPC resources of the SDumont supercomputer, which has contributed to the research results reported within this paper. URL: \url{http://sdumont.lncc.br}. This research has made use of the CTAO instrument response functions provided by the CTAO Consortium, see \url{https://www.cta-observatory.org/science/cta-performance/} (version prod5 v0.1; \url{https://doi.org/10.5281/zenodo.5499840}) for more details. This work made use of Gammapy~\cite{gammapy:2023}, a community-developed Python package. The Gammapy team acknowledges all Gammapy past and current contributors, as well as all contributors of the main Gammapy dependency libraries: \href{https://numpy.org/}{NumPy}, \href{https://scipy.org/}{SciPy}, \href{http://www.astropy.org}{Astropy}, \href{https://astropy-regions.readthedocs.io/}{Astropy Regions}, \href{https://scikit-hep.org/iminuit/}{iminuit}, \href{https://matplotlib.org/}{Matplotlib}. We also gratefully acknowledge the use of the \href{https://naima.readthedocs.io/}{Naima} software package. This research has made use of the VizieR catalog access tool, CDS, Strasbourg, France (DOI: 10.26093/cds/vizier). The original description of the VizieR service was published in 2000, A\&AS 143, 23.

% Bibliography

%% [A] Recommended: using JHEP.bst file
\bibliographystyle{JHEP}
\bibliography{biblio.bib}

\end{document}